\documentclass[aip,graphicx,amsmath,amssymb, reprint]{revtex4-1}

\usepackage{bm}
\usepackage{physics}
\usepackage{graphicx}
\usepackage{xcolor}
\usepackage{comment}

\draft 

\begin{document}


\title{Theoretical study of orbital torque: Dependence on ferromagnet species and nonmagnetic layer thickness} 



\author{Daegeun~Jo}
\email{daegeun.jo@physics.uu.se}
\affiliation{Department of Physics and Astronomy, Uppsala University, P.O. Box 516, SE-75120 Uppsala, Sweden}
\affiliation{Wallenberg Initiative Materials Science for Sustainability, Uppsala University, SE-75120 Uppsala, Sweden}

\author{Peter~M.~Oppeneer}
\email{peter.oppeneer@physics.uu.se}
\affiliation{Department of Physics and Astronomy, Uppsala University, P.O. Box 516, SE-75120 Uppsala, Sweden}
\affiliation{Wallenberg Initiative Materials Science for Sustainability, Uppsala University, SE-75120 Uppsala, Sweden}


\date{\today}

\begin{abstract}
The manipulation of magnetization in ferromagnetic metals (FMs) through orbital torque (OT) has emerged as a promising route for energy-efficient magnetic devices without relying on heavy metals. While Ti and Cu are among the most extensively studied light nonmagnetic metals (NMs) for OT devices, theoretical calculations of the resulting torque have remained limited. Here, we present a systematic and quantitative theoretical study of current-induced torques in Ti/FM and Cu/FM (FM = Co, Ni) bilayers using realistic tight-binding models derived from \textit{ab initio} electronic structures. We find that the torque in Ti/FM is larger for Ni than for Co, but this trend does not necessarily hold in Cu/FM, revealing that the FM dependence of OT is not universal but varies with the orbital current source. Moreover, the dependence of OT on NM thickness clearly indicates its NM bulk origin in both Ti- and Cu-based systems. Notwithstanding, the quantitative characteristics of OT cannot be explained by a simplified picture based on the individual bulk properties of the NM or FM layers. These results provide microscopic insight and practical guidance for designing light-metal-based orbitronic devices.
\end{abstract}


\maketitle 


\section{Introduction}\label{sec:intro}

The manipulation of magnetization by electrical means has become a central topic in spintronics. Spin-orbit torques (SOTs)\cite{manchon2019current, han2021spin, shao2021roadmap} exploit efficient charge-to-spin conversion, typically achieved by the spin Hall effect (SHE) or the spin Rashba-Edelstein effect through the spin-orbit coupling (SOC) of a nonmagnetic metal (NM). The injected spin angular momentum exerts a torque on the magnetization of an adjacent ferromagnetic metal (FM) layer. Recently, an alternative mechanism, known as the orbital torque (OT),\cite{go2020orbital, ding2020harnessing, kim2021nontrivial, lee2021orbital, lee2021efficient, hayashi2023observation} has emerged as a new route to control magnetization through orbital angular momentum (OAM) transfer from a NM layer to a FM layer. OTs arise from an OAM current (or simply, orbital current) generated by the orbital Hall effect (OHE), or from an interfacial OAM accumulation by the orbital Rashba-Edelstein effect, both of which can occur even in the absence of SOC.~\cite{go2018intrinsic, salemi2019orbitally} For the transferred OAM to exert a torque on the magnetization of the FM layer, SOC must eventually intervene, since there is no direct exchange coupling between OAM and the local spin moment.~\cite{go2020orbital} Thus, in contrast to the SOT mechanism, in which the spin current source itself requires SOC to generate spin currents, the OT mechanism assigns the role of SOC in orbital-to-spin conversion to the layer that receives the orbital current, rather than to its source. This mechanism opens a distinct pathway to efficient torque generation using light NMs with weak SOC, drawing increasing attention to the emerging field of orbitronics.~\cite{go2021orbitronics, jo2024spintronics, ando2025orbitronics}

While a variety of light NMs have been examined for OT devices,\cite{ando2025orbitronics} Ti~\cite{hayashi2023observation, choi2023observation, zhao2025artificial, xu2025observation, shin2025enhanced, hayashi2025crystallographic, xu2025orbital, zheng2025manipulating} and Cu~\cite{an2016spin, kim2021nontrivial, ding2020harnessing, tazaki2020current, kim2023oxide, krishnia2024quantifying, ding2024mitigation, damerio2025tunable, li2025oxidation, yi2025large} are among the most extensively studied materials. Ti is a representative orbital Hall material, as experimentally confirmed through magneto-optical detection~\cite{choi2023observation} and transmission electron microscopy.~\cite{idrobo2024direct} The substantial orbital current generated by the OHE in Ti serves as a source of the OT in Ti/FM bilayers. Although the OHE itself does not require SOC, the OT mechanism relies on orbital-to-spin conversion mediated by SOC. To enhance this conversion efficiency, one typically employs a magnetic layer with stronger SOC~\cite{xu2025orbital} or inserts a thin heavy metal layer between the Ti and FM layers.\cite{xu2025observation, shin2025enhanced, zheng2025manipulating} In this regard, a pronounced dependence on the FM species is a hallmark feature of OT. Notably, even among the 3$d$ FMs, Ni has been reported to generate larger OTs than other 3$d$ FMs, which is attributed to its slightly more efficient orbital-to-spin conversion.~\cite{go2020theory, lee2021orbital, lee2021efficient, dutta2022observation, bose2023detection, fukunaga2023orbital, moriya2024observation} This superiority of Ni has also been observed in Ti/FM bilayers.~\cite{hayashi2023observation, hayashi2025crystallographic} Another remarkable feature of Ti-based OT devices is the long characteristic length in the Ti-thickness dependence, reaching several tens of nanometers.~\cite{choi2023observation, hayashi2023observation, xu2025observation, hayashi2025crystallographic} Although such long-range behavior is not yet fully understood, it indicates that the bulk Ti layer serves as the primary source of the OT.

On the other hand, for Cu-based OT devices, most studies have employed the orbital Rashba-Edelstein effect activated by oxidation of the Cu surface,~\cite{kim2021nontrivial, ding2020harnessing, tazaki2020current, kim2023oxide, krishnia2024quantifying, ding2024mitigation, damerio2025tunable, li2025oxidation, yi2025large} rather than the OHE of bulk Cu. In these oxidized Cu/FM structures, enhancement of OT by inserting a heavy metal layer,\cite{ding2020harnessing, krishnia2024quantifying, li2025oxidation} as well as a dependence on the FM species, has been also observed, similar to the behavior in Ti/FM devices. For example, an earlier experiment~\cite{kim2021nontrivial} reported that the OT is significantly larger when CoFe or Fe is used as the FM, while it becomes negligible for Ni and permalloy (Py). This FM dependence appears inconsistent with the SOC-based interpretation established for other OT systems.~\cite{go2020theory, hayashi2023observation, hayashi2025crystallographic, lee2021orbital, lee2021efficient, dutta2022observation, bose2023detection, fukunaga2023orbital, moriya2024observation} This seeming discrepancy can be reconciled by attributing the pronounced FM dependence to variations in atomic intermixing at the interface,~\cite{kim2021nontrivial} whose influence outweighs that of differences in intrinsic SOC. Indeed, large OTs have been reported for both CoFe~\cite{kim2023oxide, ding2024mitigation} and Py.~\cite{an2016spin, okano2019nonreciprocal, tazaki2020current, damerio2025tunable} Such diverse FM-dependent behavior raises a key question: to what extent does SOC influence the OT among different 3$d$ FMs? 

Addressing this question calls for a comparative theoretical study of OTs in NM/FM bilayers with various combinations of NMs and FMs. Given the long-range behavior of OTs,~\cite{go2023long, choi2023observation, hayashi2023observation, xu2025observation, hayashi2025crystallographic, bose2023detection, fukunaga2023orbital, moriya2024observation} it is desirable to investigate sufficiently large bilayer systems, which poses a challenge for fully \textit{ab initio} methods. Although several theoretical studies~\cite{go2020theory, salemi2021quantitative, go2023long, nikolaev2024large, devda2025anatomy, pezo2025anatomy} have examined the OT in various heterostructures, a systematic theoretical investigation of the FM dependence of OT in light NM/FM bilayers has, to our knowledge, not yet been performed. It is worth noting that the FM dependence could also originate from a self-induced SOT within the FM layer, also known as the anomalous SOT.~\cite{wang2019anomalous} Examining the dependence of the torque on material combination and layer thickness will provide valuable insight into distinguishing  OT from other possible contributions.

In this work, we perform a systematic theoretical study of current-induced torques in NM/FM bilayers, where NM = Ti or Cu and FM = Co or Ni. By employing realistic tight-binding models with parameters extracted from \textit{ab initio} calculations, we achieve relatively large system sizes---up to 50 atomic layers ($\gtrsim 10$~nm). For Ti/FM bilayers, we find that the torque is larger for Ni, consistent with experimental observations.~\cite{hayashi2023observation, hayashi2025crystallographic} For Cu-based systems, we focus on pristine Cu/FM bilayers without oxidation. Although OT in these systems is generally assumed to be negligible, our results reveal non-negligible OT signals, indicating a finite OHE in Cu as predicted theoretically~\cite{jo2018gigantic, salemi2022first, go2024first} and observed experimentally.~\cite{rothschild2022generation, ben2025measurement} Interestingly, the trend of FM dependence reverses in Cu/FM systems under certain conditions, with Co producing a stronger torque. This finding demonstrates that the FM dependence of OT is not universal but can vary depending on the orbital current source, even when only intrinsic effects are considered. We also examine the NM thickness dependence, confirming the bulk NM origin of OTs in both Ti/FM and Cu/FM systems. In Cu/Ni, the bulk Cu contribution exhibits the opposite sign to that in Cu/Co, cautioning that OT cannot be directly predicted from the bulk orbital Hall conductivity of the NM. Consequently, our results demonstrate that the characteristics of OT cannot be explained solely by the individual bulk properties of the NM or FM layers, but depend critically on their material combination. These findings provide microscopic insight into the design of efficient orbitronic devices.

\section{Theoretical Methodology}\label{sec:method}

\subsection{Tight-binding models based on \textit{ab initio} electronic structures}\label{sec:method_tight}

We construct realistic tight-binding models for NM/FM systems using the Wannier functions extracted from \textit{ab initio} band structures. The bulk band structures of the NMs (Ti and Cu) and FMs (Co and Ni) are obtained using the full-potential linearized augmented plane-wave code \texttt{FLEUR}.~\cite{fleurWeb, fleurCode} The Perdew-Burke-Ernzerhof functional~\cite{perdew1996generalized} within the generalized gradient approximation is used for the exchange-correlation functional. All systems are assumed to have the face-centered cubic (fcc) structure, and the lattice constants of the FMs are set equal to those of the corresponding NMs. Specifically, for Co and Ni, two sets of band structures are independently calculated using either the lattice constant of Ti~\cite{chakraborty2011thickness} ($a_\mathrm{Ti} = 4.16$~\r{A}) or that of Cu~\cite{straumanis1969lattice} ($a_\mathrm{Cu}= 3.61$~\r{A}), assuming that the FM layer adopts the same structure as the adjoining NM layer in each NM/FM bilayer. The magnetization of Ni and Co is aligned along the normal vector to the fcc(111) plane. SOC is treated within the second variation scheme,~\cite{li1990magnetic} and the Brillouin zone is sampled using a $12 \times 12 \times 12$ Monkhorst-Pack $\mathbf{k}$-mesh.~\cite{monkhorst1976special}

Based on the \textit{ab initio} electronic structures, we use the \texttt{WANNIER90} code~\cite{pizzi2020wannier90} to construct Wannier functions for each system. Instead of maximally localized Wannier functions, we obtain a set of 18 projected Wannier functions corresponding to the $s$, $p_x$, $p_y$, $p_z$, $d_{xy}$, $d_{yz}$, $d_{zx}$, $d_{z^2}$, and $d_{x^2-y^2}$ orbitals for each spin, which yield reasonable accuracy for simple metallic systems.~\cite{freimuth2008maximally} A uniform $8 \times 8 \times 8$ $\mathbf{k}$-mesh is used for sampling the Brillouin zone. During the disentanglement procedure, the frozen energy window is defined from the bottom of the valence bands to 5~eV above the Fermi energy. The obtained Wannier function is denoted as $ \vert  n \mathbf{R} \rangle $, where $n$ is the band (orbital) index and $\mathbf{R}$ is the lattice vector. The matrix elements of the Hamiltonian $\hat{H}$ and spin $\hat{\mathbf{S}}$ operators, evaluated in the Bloch basis, are transformed into the Wannier representation as $ \langle m \mathbf{0} \vert \hat{H} \vert n \mathbf{R} \rangle $ and $ \langle m \mathbf{0} \vert  \hat{\mathbf{S}} \vert n \mathbf{R} \rangle $, respectively. The Hamiltonian operator can be decomposed as $\hat{H} = \hat{H}_0 + \hat{H}_\mathrm{XC} + \hat{H}_\mathrm{SO}$, where $\hat{H}_0$ contains the kinetic and potential energies, $\hat{H}_\mathrm{XC}$ represents the exchange coupling (for Co and Ni), and $\hat{H}_\mathrm{SO}$ accounts for SOC. These matrix elements are then used as input parameters for constructing the tight-binding models of the NM/FM bilayers, as described below.

\begin{figure}[t]
	\center\includegraphics[width=0.48\textwidth]{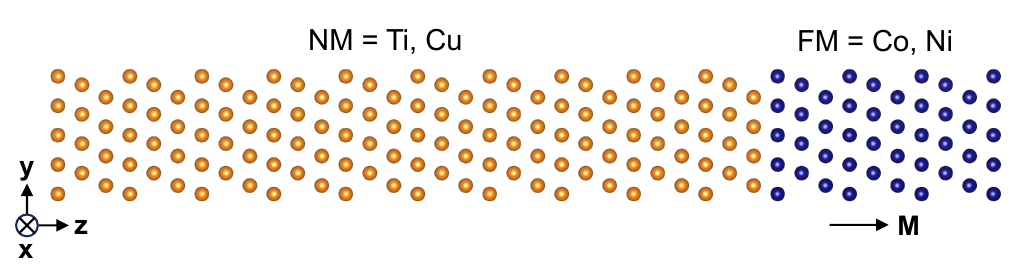}
	\caption{Illustration of the tight-binding model for the NM/FM bilayer structure with fcc(111) orientation. The presented system is drawn for $N_\mathrm{NM}=30$ and $N_\mathrm{FM}=10$. The system is periodic along $x$ and $y$ directions. The FM layer has an out-of-plane magnetization $\mathbf{M}$ along $\hat{\mathbf{z}}$.}
	\label{fig:structure} 
\end{figure}

The NM/FM bilayer structures are assumed to be stacked along the (111) direction of the fcc lattice. Each system consists of $N_\mathrm{NM}$ atomic layers of NM and $N_\mathrm{FM}$ atomic layers of FM, as illustrated in Fig.~\ref{fig:structure}. The NM/FM bilayer is modeled as a two-dimensional hexagonal lattice with a lattice constant of $a_\mathrm{NM} / \sqrt{2}$ and an interlayer spacing of $a_\mathrm{NM} / \sqrt{3}$, where $a_\mathrm{NM}$ ($= a_\mathrm{Ti}$ or $a_\mathrm{Cu}$) is the bulk lattice constant of the corresponding NM. Based on the bulk Wannier functions $\vert n \mathbf{R} \rangle$, we construct the operators $ \hat{\mathcal{O}}_\mathrm{film} $ for the NM/FM film structure, corresponding to an arbitrary bulk operator $\hat{\mathcal{O}}$, by introducing a set of $18 \times (N_\mathrm{NM} + N_\mathrm{FM})$ Wannier functions $ \vert n_i \mathbf{R}_\mathrm{2D} \rangle $. Here, $\mathbf{R}_\mathrm{2D}$ is a two-dimensional lattice vector, and $n_i$ denotes the $n$ orbital in the $i$-th layer. The Wannier center $\mathbf{r}_{n_i}$ is assumed to coincide with the atomic position $\mathbf{r}_i$ of the $i$-th layer. The matrix elements $  \langle m_j \mathbf{0} \vert \hat{\mathcal{O}}_\mathrm{film} \vert n_i \mathbf{R}_\mathrm{2D} \rangle $, describing hopping between layers $i$ and $j$, are derived from the corresponding bulk matrix elements $ \langle m \mathbf{0} \vert \hat{\mathcal{O}} \vert n \mathbf{R} \rangle $ whenever the condition $\mathbf{R}_\mathrm{2D} + \mathbf{r}_{i} - \mathbf{r}_{j} = \mathbf{R} $ is satisfied. When both layers $i$ and $j$ correspond to the same atomic species, the matrix element in the bilayer is taken to be identical to that of the bulk system. For matrix elements connecting different atomic species, such as hopping terms between NM and FM layers, $  \langle m_j \mathbf{0}  \vert \hat{\mathcal{O}}_\mathrm{film} \vert n_i \mathbf{R}_\mathrm{2D} \rangle $ is approximated by the average of the bulk values of the two materials. Hereafter, we omit the subscripts in $  \langle m_j \mathbf{0} \vert \hat{\mathcal{O}}_\mathrm{film} \vert n_i \mathbf{R}_\mathrm{2D} \rangle $ and simply write $  \langle m \mathbf{0} \vert \hat{\mathcal{O}} \vert n \mathbf{R} \rangle $ to represent matrix elements for the bilayer system. For the position operator $\hat{\mathbf{r}}$, we define $\langle m \mathbf{0} \vert \hat{\mathbf{r}} \vert n \mathbf{R} \rangle = \mathbf{r}_n \delta_{nm} \delta_{0,\mathbf{R}}$, where $\delta_{nm}$ is the Kronecker delta.

Once the matrix elements $\bra{m \mathbf{0}} \hat{\mathcal{O}} \ket{n \mathbf{R}}$ are obtained ($\hat{\mathcal{O}} = \hat{H}, \hat{\mathbf{S}}, \hat{\mathbf{r}}$), they are interpolated onto a dense $150 \times 150$ $\mathbf{k}$-mesh in momentum space. The Bloch-like states in the Wannier gauge are defined as
	\begin{equation}
		\ket{\psi_{n\mathbf{k}}^\mathrm{W}} = \frac{1}{\sqrt{N}} \sum_\mathbf{R} e^{i \mathbf{k} \cdot \mathbf{R}} \ket{n \mathbf{R}} ,
	\end{equation}
	where $N$ is the number of lattice vectors. The corresponding operator matrix elements in the Wannier gauge are then given by
	\begin{equation}
		\mathcal{O}_{mn}^\mathrm{W}(\mathbf{k}) = \bra{\psi_{m\mathbf{k}}^\mathrm{W}} \hat{\mathcal{O}} \ket{\psi_{n\mathbf{k}}^\mathrm{W}} = \frac{1}{N} \sum_\mathbf{R} e^{i \mathbf{k} \cdot \mathbf{R}} \bra{m \mathbf{0}} \hat{\mathcal{O}} \ket{n \mathbf{R}} .
	\end{equation}
    Since $\ket{\psi_{n\mathbf{k}}^\mathrm{W}}$ is not an eigenstate of the Hamiltonian, the eigenstates $\ket{\psi_{n\mathbf{k}}}$ and eigenvalues $\epsilon_{n\mathbf{k}}$, satisfying $\hat{\mathcal{H}} \ket{\psi_{n\mathbf{k}}} = \epsilon_{n\mathbf{k}} \ket{\psi_{n\mathbf{k}}}$, are obtained by diagonalizing $\hat{\mathcal{H}}^\mathrm{W}(\mathbf{k})$. The resulting matrix elements  $\bra{\psi_{m\mathbf{k}}} \hat{\mathcal{O}} \ket{\psi_{n\mathbf{k}}}$ in the Hamiltonian gauge are then used in the linear-response calculation, as described in the following subsection.

	\subsection{Theoretical description of the torque}\label{sec:method_torque}
	To study OT, we employ the continuity equation for angular momentum transport in a general magnetic system described by the Hamiltonian $\hat{H} = \hat{H}_0 + \hat{H}_\mathrm{XC} + \hat{H}_\mathrm{SO}$. By combining the two equations of motion for spin and OAM, which are coupled to each other via SOC, one obtains the continuity equation for the total angular momentum operator $\hat{\mathbf{J}} = \hat{\mathbf{L}} + \hat{\mathbf{S}}$ as follows:~\cite{haney2010current, go2020theory}
	\begin{equation}\label{eq:continuity}
		\frac{\partial \hat{J}_i (\mathbf{r})}{\partial t} = - \nabla \cdot \hat{\mathbf{Q}}^{J_i} (\mathbf{r}) + \hat{T}_i^\mathrm{XC}(\mathbf{r}) + \hat{T}_i^\mathrm{CF}(\mathbf{r}).
	\end{equation}
	Here, $\hat{\mathcal{O}}(\mathbf{r}) = \hat{\psi}^\dagger(\mathbf{r}) \hat{\mathcal{O}} \hat{\psi}(\mathbf{r})$ denotes the local density operator for an arbitrary single-particle operator $\hat{\mathcal{O}}$, where $\hat{\psi}^\dagger(\mathbf{r})$ and $\hat{\psi}(\mathbf{r})$ are the creation and annihilation operators at position $\mathbf{r}$, respectively, $\hat{\mathbf{Q}}^{\mathbf{J}} = \frac{1}{2} (\hat{\mathbf{v}} \hat{\mathbf{J}} + \hat{\mathbf{J}} \hat{\mathbf{v}} ) $ is the total angular momentum current operator, and
	\begin{align}
	\hat{\mathbf{T}}^\mathrm{XC} & = \frac{i}{\hbar} [  \hat{H}_\mathrm{XC}, \hat{\mathbf{S}} ], \label{eq:t_xc} \\
	\hat{\mathbf{T}}^\mathrm{CF} & = \frac{i}{\hbar} [ \hat{H}_\mathrm{0}, \hat{\mathbf{L}} ] \label{eq:t_cf}
	\end{align}
	are the exchange torque and the crystal field torque operators, respectively. Note that $\hat{H}_\mathrm{0}$ in Eq.~\eqref{eq:t_cf} contains the rotationally invariant kinetic energy ($\frac{-\hbar^2\nabla^2}{2m}$) and the crystal potential energy, but only the latter does not commute with $\hat{\mathbf{L}}$ and therefore contributes to $\hat{\mathbf{T}}^\mathrm{CF}$.

	Equation~\eqref{eq:continuity} relates the local accumulation rate of angular momentum to the divergence (or flux) of the angular momentum current, e.g., generated by orbital and spin Hall effects. The torque terms act as sources or sinks that violate the conservation of total angular momentum in an electronic system, describing the transfer of angular momentum from conduction electrons to other degrees of freedom. The exchange torque $\hat{\mathbf{T}}^\mathrm{XC}$ describes transfer of angular momentum to the local magnetic moment via exchange coupling. The magnetization dynamics probed in torque experiments (including SOT, OT, etc.) is captured by this term. Therefore, we mainly focus on $\hat{\mathbf{T}}^\mathrm{XC}$ throughout this paper. The crystal field torque $\hat{\mathbf{T}}^\mathrm{CF}$ describes transfer of angular momentum to the lattice via the crystal field, i.e., quenching of OAM. Although this term is often disregarded, Eq.~\eqref{eq:continuity} explicitly shows that it plays an important role in angular momentum transport by connecting the non-conserved angular momentum current $\hat{\mathbf{Q}}^{\mathbf{J}}$ to experimentally relevant observables such as $\hat{\mathbf{J}}$ and $\hat{\mathbf{T}}^\mathrm{XC}$. Thus, the conventional definition of angular momentum current~\cite{tanaka2008intrinsic, kontani2009giant, go2018intrinsic, jo2018gigantic, salemi2022first} alone is insufficient to determine the angular momentum accumulation or the torque. Because our approach directly computes $\hat{\mathbf{T}}^\mathrm{XC}$ exerted on the magnetization, this issue associated with defining the angular momentum current does not affect our analysis. In the presence of SOC, a complete separation of spin and orbital contributions to $\hat{\mathbf{T}}^\mathrm{XC}$ is generally nontrivial. Nonetheless, we will demonstrate that OT is the dominant mechanism in our systems.

\subsection{Linear-response calculation of the current-induced torque}\label{sec:method_linear}
	
The current-induced torques in NM/FM bilayers are calculated within the framework of linear-response theory. Under an in-plane electric field $\mathbf{E} =E_x \hat{\mathbf{x}}$, the induced torque generally contains two components: a field-like torque $\propto \hat{\mathbf{M}} \times \hat{\mathbf{y}}$ and a damping-like torque $\propto \hat{\mathbf{M}} \times (\hat{\mathbf{M}} \times \hat{\mathbf{y}})$, where $\hat{\mathbf{M}}$ is the unit vector of magnetization.~\cite{manchon2019current} For an out-of-plane magnetization in the FM, i.e., $\hat{\mathbf{M}} = \hat{\mathbf{z}}$, the field-like and damping-like torques are directed along $\hat{\mathbf{x}}$ and $\hat{\mathbf{y}}$, respectively. In this work, we focus on the damping-like torque, which has been primarily investigated in orbitronics~\cite{ando2025orbitronics} and is known to play a crucial role in SOT device functionalities.~\cite{shao2021roadmap} 

Within linear-response theory, the torque $\mathbf{T}$ induced by the electric field $\mathbf{E}$ is expressed as 
\begin{equation}\label{eq:torque}
    T_{i} = t_{ij} E_j,
\end{equation}
where $t_{ij}$ is the component of the linear-response tensor $\mathbf{t}$, referred to as the torkance, and $i$ and $j$ denote Cartesian components. In the Kubo formalism, $t_{ij}$ is decomposed into intraband and interband contributions, $t_{ij} = t_{ij}^\mathrm{intra} + t_{ij}^\mathrm{inter}$, each of which is evaluated as follows:~\cite{freimuth2014spin, go2020theory}
\begin{eqnarray}
	t_{ij}^\mathrm{intra} = &&
	\frac{e\hbar}{\Gamma} \int \frac{d^2\mathbf{k}}{(2\pi)^2} \sum_{n} 
	(\frac{\partial f_{n\mathbf{k}}}{\partial \epsilon_{n\mathbf{k}}} ) \nonumber \\ 
	&& \times \langle \psi_{n\mathbf{k}} \vert \hat{T}_i^\mathrm{XC}  \vert \psi_ {n\mathbf{k}}\rangle 
		\langle \psi_{n\mathbf{k}} \vert \hat{v}_j \vert \psi_{n\mathbf{k}} \rangle ,  \label{eq:t_ij_intra} \\
	t_{ij}^\mathrm{inter} = &&
	-e \hbar \int \frac{d^2\mathbf{k}}{(2\pi)^2} \sum_{n \neq m} 
	(f_{n\mathbf{k}} - f_{m\mathbf{k}}) \nonumber \\
  && \times \mathrm{Im}
	\left[
	\frac
	{\langle \psi_{n\mathbf{k}} \vert \hat{T}_i^\mathrm{XC}  \vert \psi_ {m\mathbf{k}}\rangle 
		\langle \psi_{m\mathbf{k}} \vert \hat{v}_j \vert \psi_{n\mathbf{k}} \rangle 
	}
	{(\epsilon_{n\mathbf{k}} - \epsilon_{m\mathbf{k}})( \epsilon_{n\mathbf{k}} - \epsilon_{m\mathbf{k}} + i\Gamma )} 
	\right], \label{eq:t_ij_inter}
\end{eqnarray}
where $e>0$ is the elementary charge, $f_{n\mathbf{k}} = 1/[e^{ (\epsilon_{n\mathbf{k}} - \mu ) / k_\mathrm{B}T} + 1 ]$ is the Fermi-Dirac distribution function, with the chemical potential $\mu$ and $k_\mathrm{B}T = 25$~meV, $\hat{\mathbf{v}} = \frac{i}{\hbar}[\hat{H},\hat{\mathbf{r}}]$ is the velocity operator, and $\hat{\mathbf{T}}^\mathrm{XC}$ is the exchange torque operator defined by Eq.~\eqref{eq:t_xc}. The effect of disorder is approximately described by a constant energy broadening $\Gamma$. The OAM current or accumulation is not explicitly considered in the torque calculations using Eqs.~\eqref{eq:t_ij_intra} and \eqref{eq:t_ij_inter}, but its contribution to the OT is implicitly included (see Sec.~\ref{sec:method_torque}). The intraband term $t_{ij}^\mathrm{intra}$ in Eq.~\eqref{eq:t_ij_intra} corresponds to the (Drude-like) Boltzmann contribution, which scales with the momentum relaxation time $\tau \propto 1/\Gamma$. In contrast, the interband term $t_{ij}^\mathrm{inter}$ in Eq.~\eqref{eq:t_ij_inter} arises from the Berry curvature mechanism and represents an intrinsic contribution. Notably, $t_{ij}^\mathrm{intra}$ and $t_{ij}^\mathrm{inter}$ are odd and even under time-reversal, respectively. Symmetry considerations indicate that $t_{ij}^\mathrm{intra}$ gives rise to the field-like torque, whereas $t_{ij}^\mathrm{inter}$ leads to the damping-like torque.~\cite{go2020theory} Therefore, we calculate only $t_{yx}^\mathrm{inter} \equiv t_{yx}$ to describe the damping-like torque induced by an electric field applied along $\hat{\mathbf{x}}$.

\section{Results and Discussion}\label{sec:results}

\subsection{Phenomenological mechanism}\label{sec:phenomenology}

Before presenting the numerical results, we first discuss the phenomenology of the current-induced torques in NM/FM bilayers. It is important to note that our calculations yield the total current-induced torque acting on the magnetization, rather than the OT contribution alone. While the damping-like torque evaluated by Eq.~\eqref{eq:t_ij_inter} originates from intrinsic mechanisms, multiple contributions may coexist, including OT, SOT, and self-induced torque within the FM layer.~\cite{go2020theory} For Ti/FM and Cu/FM bilayers, however, the SOT contribution can be safely neglected since both Ti and Cu exhibit negligible SHEs.\cite{jo2018gigantic, salemi2022first, go2024first} Therefore, our calculations mainly capture the OT or self-induced torque components. Distinguishing these contributions unambiguously is generally nontrivial, both theoretically and experimentally. Below, we discuss the physical mechanisms underlying these two torques and their expected dependences on the FM species and NM thickness.

\begin{figure}[t]
	\center\includegraphics[width=0.48\textwidth]{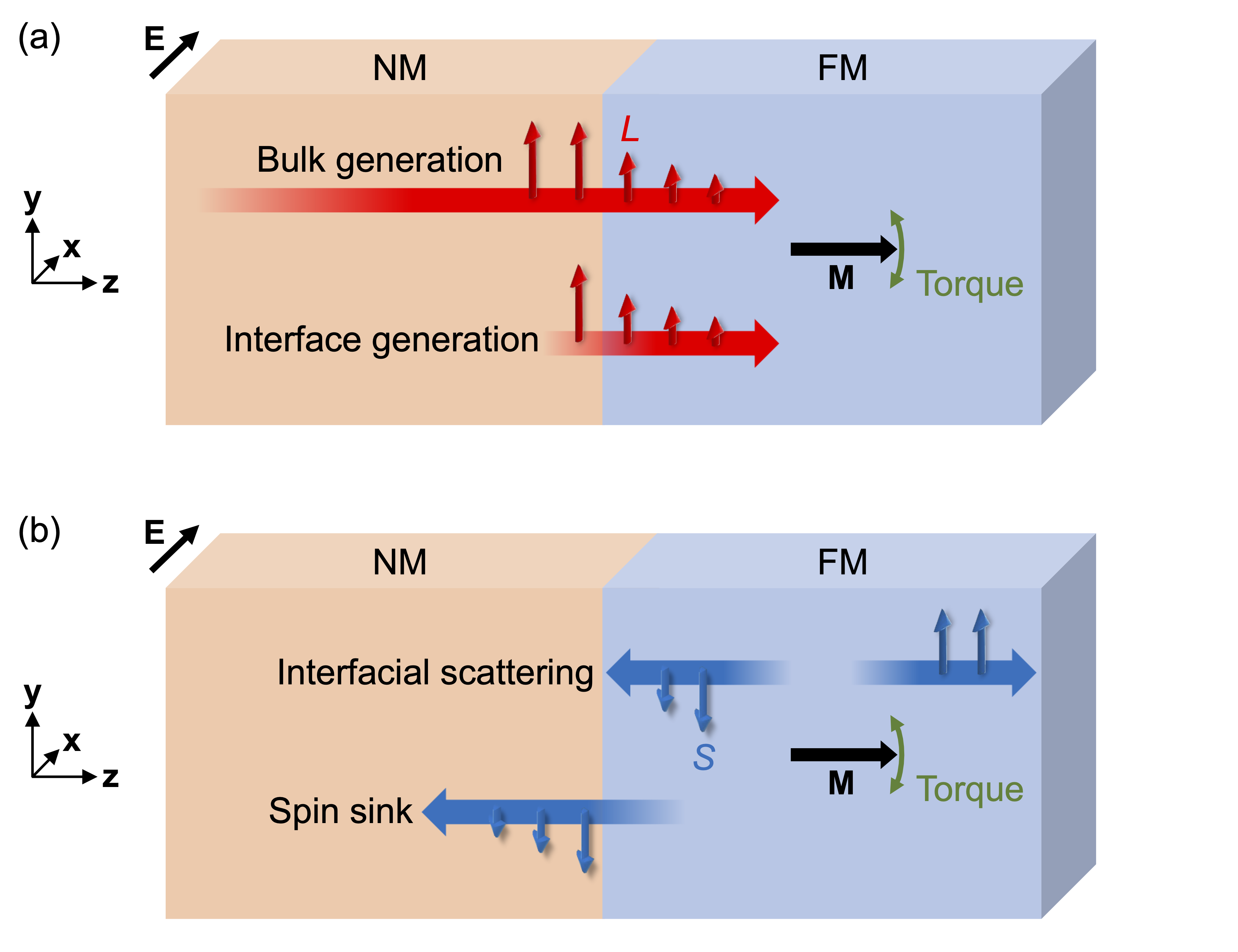}
	\caption{Schematic illustration of the (a) OT and (b) self-induced SOT in a NM/FM bilayer. The red and blue arrows represent the OAM ($L$) and spin ($S$), respectively. For each case, two possible contributions associated with the bulk NM and the NM/FM interface are shown.}
	\label{fig:schematic_torque} 
\end{figure}

Let us consider a light NM/FM bilayer structure, where the SOC in the NM is negligible, such that it creates only orbital currents but not spin currents. The FM is assumed to possess weak SOC that nevertheless allows moderate orbital-to-spin conversion. We confine our discussion to intrinsic effects governed by the electronic band structure and disregard extrinsic factors such as crystallinity or interfacial quality. Figure~\ref{fig:schematic_torque}(a) illustrates how an orbital current generated in the NM gives rise to an OT on the FM layer.~\cite{go2020orbital} When the orbital current transmits across the NM/FM interface, it undergoes intrinsic relaxation and is absorbed into the lattice and spin degrees of freedom via the crystal field and SOC, respectively. Even if a significant portion of the OAM is quenched by the crystal field, the remaining OAM can still lead to a substantial torque on the FM layer. The torque efficiency is then determined by how effectively the OAM couples to the local spin moment through SOC and exchange interaction, competing with quenching due to the crystal field. Therefore, the OT mechanism inherently depends on the FM's intrinsic properties, particularly its SOC, which governs the efficiency of orbital-to-spin conversion. Additional factors such as interface transparency~\cite{lyalin2024interface} or orbital texture matching between the NM and FM layers~\cite{hayashi2025crystallographic} may also play a role, though their effects are not yet fully understood.

Meanwhile, the orbital current may originate either from the bulk or from a confined interfacial region. If it is generated only near the interface, its contribution to the OT should be largely independent of NM thickness. In contrast, if the orbital current is generated in the NM bulk, the transferred OAM, and thus the OT, should increase with increasing NM thickness, up to a characteristic length scale associated with OAM generation in the NM layer. This length scale is not necessarily identical to the orbital diffusion length. Considering that both the intrinsic OHE~\cite{go2018intrinsic} and the damping-like torque described by Eq.~\eqref{eq:t_ij_inter} arise from interband coherence in response to the electric field, the thickness dependence may also reflect an interband coherence length. The relevant characteristic length may therefore vary with the regime of OAM transport, which remains under active debate.~\cite{rang2024orbital, sohn2024dyakonov, tang2024role, barbosa2025extrinsic, valet2025quantum}

We now turn to the self-induced SOT, illustrated in Fig.~\ref{fig:schematic_torque}(b). We consider a situation in which the bulk FM generates a spin current via the SHE, leading to spin accumulation within the FM layer itself. In an inversion symmetric geometry---e.g., a FM layer sandwiched between identical top and bottom contacts---the spin accumulations at the two interfaces have the same magnitude with opposite signs, so the resulting torque cancels out. In a NM/FM bilayer, however, the lack of inversion symmetry causes the spin accumulation at the NM/FM interface and at the opposite FM surface to differ, producing a net SOT. Because the spin Hall conductivity varies substantially across 3$d$ FMs,\cite{amin2019intrinsic, salemi2022first, go2024first} this self-induced SOT also strongly depends on the choice of FM.~\cite{wang2019anomalous} A recent experiment~\cite{liu2025absence} reported that the FM dependence of the current-induced torque in Ta/FM bilayers originates predominantly from such a self-induced SOT, rather than the OT. It should be noted, however, that the self-induced SOT, despite its self-generating origin, depends critically not only on the material combination but also on the NM thickness, as discussed below.

The imbalance in spin accumulation can be induced by two different mechanisms. The first is an interfacial contribution, where inequivalent interfaces lead to different local spin accumulations at the top and bottom surfaces of the FM, irrespective of the bulk NM region. Such asymmetry can result from variations in interfacial spin scattering at the NM/FM interface,~\cite{amin202interfacial} which are typically material-specific. The resulting torque is expected to be largely independent of the NM thickness. The second is a bulk contribution, where the NM acts as a spin sink. When the spin current generated within the FM layer is injected into the NM layer, it experiences relaxation via the SOC of the NM bulk. The greater the spin absorption by the NM, the larger the resulting torque. Consequently, this contribution increases with NM thickness, where the relevant characteristic length is determined by the spin relaxation length of the NM. However, the spin relaxation length is typically significantly long in light NMs due to weak SOC, which implies that they are inefficient spin sinks.~\cite{kim2024spin} If the NM layer is thinner than its spin diffusion length, the spin current will be only weakly absorbed. Therefore, in nanoscale NM/FM bilayers, the self-induced SOT is expected to be much weaker for light NMs than for heavy NMs such as Ta.~\cite{liu2025absence}

Finally, we note that, in addition to the SHE, a considerable OHE can also occur in 3$d$ FMs.~\cite{salemi2022first, go2024first} This suggests the possibility of a self-induced OT arising through a similar mechanism, which, to the best of our knowledge, has not been reported before. In that case, the NM-thickness dependence of the torque would be governed by the orbital relaxation length, instead of the spin relaxation length. Importantly, this length scale is likely different from that of the OT, since the orbital sink mechanism would involve diffusive OAM relaxation rather than its coherent generation. Similarly, a distinction between the characteristic length scales for Ti measured in OT experiments (where the orbital current is injected from the NM to the FM) and in orbital pumping experiments (where the orbital current flows in the opposite direction) was recently discussed.~\cite{hayashi2024observation}

\subsection{Current-induced torques in Ti/FM and Cu/FM}\label{sec:results_torque}

\begin{figure}[t]
	\center\includegraphics[width=0.48\textwidth]{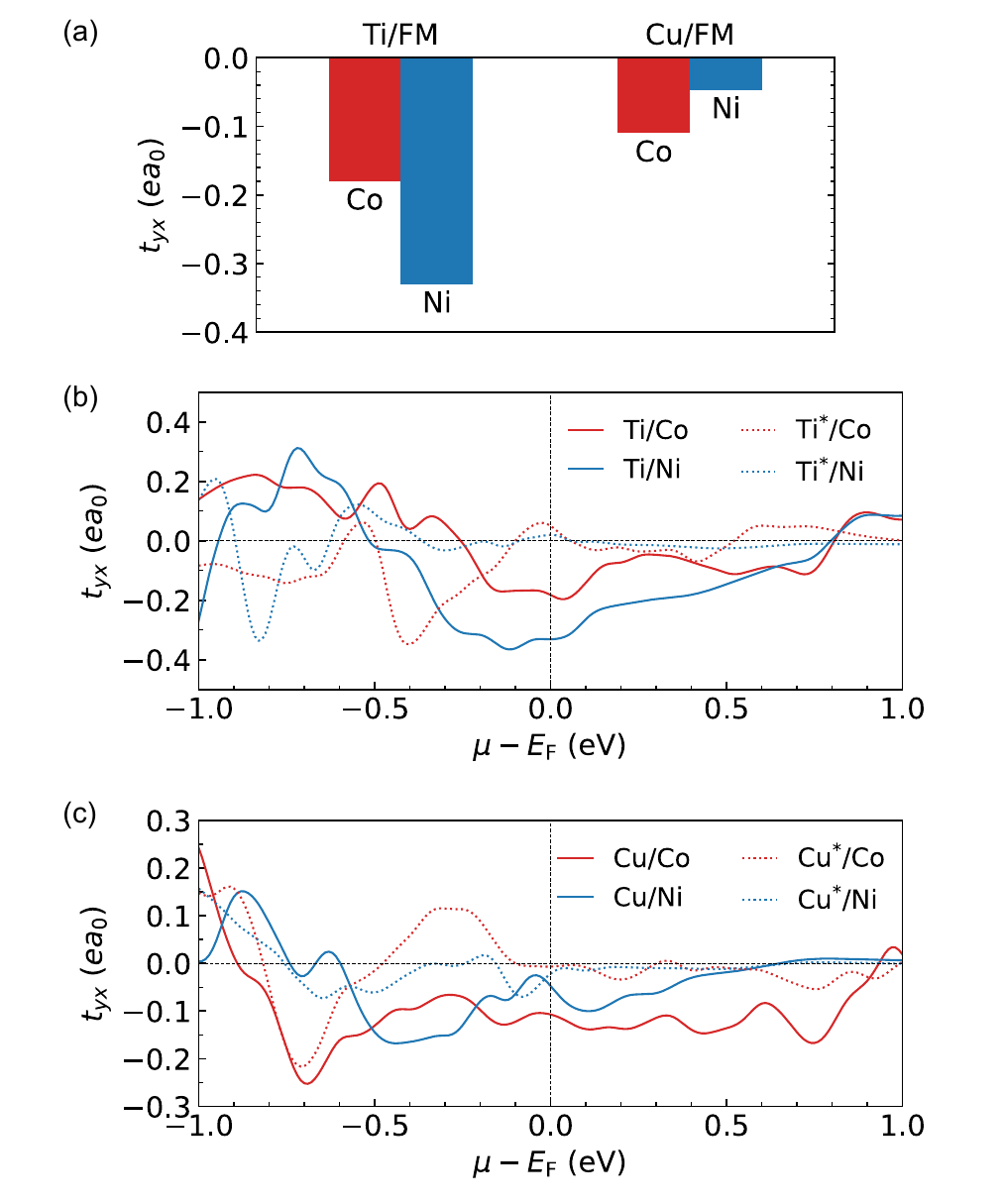}
	\caption{(a) Torkance $t_{yx}$ in Ti/FM and Cu/FM bilayers (FM = Co, Ni) with $N_\mathrm{NM}=20$ and $N_\mathrm{FM}=10$. Chemical potential dependence of $t_{yx}$ in (b) Ti/FM and (c) Cu/FM bilayers. Dotted lines represent the torque in hypothetical systems where the chemical potential of the NM layer is shifted so that the OHE is suppressed. The energy broadening is assumed to be $\Gamma = 25$~meV.}
	\label{fig:torque} 
\end{figure}

Figure~\ref{fig:torque}(a) presents the damping-like component of the torkance $t_{yx} = t_{yx}^\mathrm{inter}$ evaluated from Eq.~\eqref{eq:t_ij_inter} for Ti/FM and Cu/FM bilayers, where FM is either Co or Ni. The layer numbers are set to $N_\mathrm{NM}=20$ and $N_\mathrm{FM}=10$. All systems exhibit a negative $t_{yx}$, corresponding to the sign of the current-induced torque expected for NMs with positive spin or orbital Hall conductivity.\cite{freimuth2014spin, go2020theory} The sign of $t_{yx}$ is thus consistent with the positive orbital Hall conductivities of Ti and Cu reported in earlier theoretical works,\cite{jo2018gigantic, salemi2022first} although a more recent study predicted a negative orbital Hall conductivity for Cu.~\cite{go2024first} However, caution is warranted, as the theoretical evaluation of orbital Hall conductivity is not yet fully established. The conventional definition of orbital Hall conductivity\cite{tanaka2008intrinsic, kontani2009giant, go2018intrinsic, jo2018gigantic, salemi2022first} is based on the conventional orbital current operator $\hat{Q}_i^{L_j} = \frac{1}{2}(\hat{v}_i \hat{L}_j + \hat{L}_j \hat{v}_i )$. This definition often provides qualitatively reasonable agreement with experiments,~\cite{costa2025revealing, santos2025probing} but its calculated value varies considerably across different theoretical approaches.~\cite{salemi2022first, pezo2022orbital, go2024first, rang2025orbital} Moreover, the definition of $\hat{Q}_i^{L_j}$ itself possesses an inherent limitation stemming from the non-conservation of OAM in crystals,~\cite{choi2023observation, atencia2024intrinsic, sun2025nonconserved} as discussed in Sec.~\ref{sec:method_torque}. The orbital current undergoes a significant change due to sources such as the crystal field. Establishing a link between the conventional orbital current and experimentally measurable quantities thus requires accounting for the non-conserved part of OAM [see, e.g., Eq.~\eqref{eq:continuity}]. Our approach circumvents these issues by directly computing the well-defined observable, the torque exerted on the magnetization, without explicitly invoking OAM.

For Ti/FM bilayers, we obtain $t_{yx} = -0.18 \ ea_0$ for Ti/Co and $t_{yx} = -0.33 \ ea_0$ for Ti/Ni. Both values are notably large, comparable to those calculated for conventional SOT systems employing heavy NMs such as Pt and W.~\cite{freimuth2014spin, go2020theory} When Ti is replaced by Cu, the torque magnitude is reduced but remains finite: $t_{yx} = -0.11 \ ea_0$ for Cu/Co and $t_{yx} = -0.047 \ ea_0$ for Cu/Ni. This indicates that, even without oxidation, Cu can intrinsically generate a measurable torque. The result aligns with theoretical predictions that Cu hosts a finite OHE despite its predominantly $s$-orbital character,~\cite{jo2018gigantic, salemi2022first, go2024first} and with experimental reports of OHE in Cu,~\cite{rothschild2022generation, ben2025measurement} though a quantitative comparison remains difficult.

A key finding of this work is that the FM dependence of the OT in NM/FM bilayers varies depending on the NM species. For Ti/FM, Ni yields a larger torque than Co, which is consistent with experimentally observed large torques in Ti/Ni devices~\cite{hayashi2023observation, choi2023observation, hayashi2025crystallographic} and with the established understanding that Ni's stronger SOC enables more efficient orbital-to-spin conversion. This trend persists across a broad chemical potential range near the Fermi level, as shown in Fig.~\ref{fig:torque}(b). Although it is impractical to drastically tune the chemical potential in metallic systems, examining the energy dependence can still provide useful insights. Within the OT mechanism, the sign reversal of $t_{yx}$ with decreasing energy reflects the energy-dependent spin-orbit correlation of the FM.~\cite{go2020orbital, lee2021orbital} The peak near $-0.35$ eV~for Ti/Co is slightly shifted from the peak near $-0.6$~eV for Ti/Ni because Ni has one additional $d$ electron. Hence, the results for Ti can be reasonably explained in terms of the FM's spin-orbit correlation.

In contrast, for Cu/FM bilayers, the torque is larger for Co than Ni. Moreover, the energy dependence of $t_{yx}$ in Fig.~\ref{fig:torque}(c) does not show a clear trend between Cu/Co and Cu/Ni. This suggests that the orbital-to-spin efficiency is not the dominant factor governing the FM dependence in these systems; instead, other factors---such as interface transparency~\cite{lyalin2024interface} or orbital texture matching~\cite{hayashi2025crystallographic}---may play more decisive roles. This interpretation is reasonable considering that Ni and Co are adjacent in the Periodic Table and thus their SOC strengths are not dramatically different. Furthermore, we note that the observed behavior cannot be attributed to self-induced SOT, since Ni exhibits a substantially larger spin Hall conductivity than Co.~\cite{salemi2022first, go2024first} 

\begin{figure}[b]
	\center\includegraphics[width=0.48\textwidth]{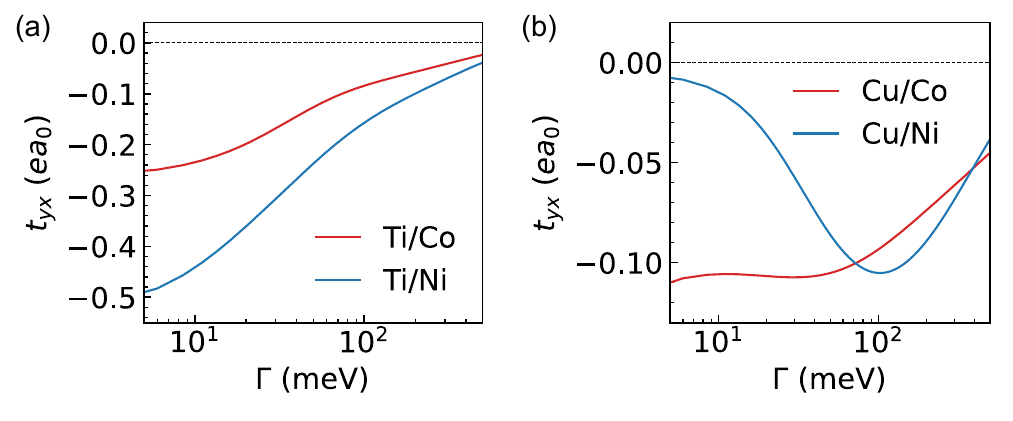}
	\caption{(a) Torkance $t_{yx}$ as a function of the broadening energy $\Gamma$ in (a) Ti/FM and (b) Cu/FM bilayers (FM = Co, Ni) with $N_\mathrm{NM}=20$ and $N_\mathrm{FM}=10$.}
	\label{fig:torque_eta} 
\end{figure}

The disorder effect is modeled by introducing a constant broadening $\Gamma$ in Eq.~\eqref{eq:t_ij_inter}. In this formalism, $\Gamma$ is associated with a finite quasiparticle lifetime, $\tau = \hbar/\Gamma$, which phenomenologically accounts for scattering in the sample. Increasing $\Gamma$ shortens $\tau$, thereby moving the system from a cleaner to a more disordered regime and making interband coherence more easily destroyed.  Figure~\ref{fig:torque_eta}(a) shows that in Ti/FM systems, $\vert t_{yx} \vert$ decreases monotonically as $\Gamma$ increases, while the relative FM dependence remains robust. The enhancement of torque when using Ni instead of Co is roughly twofold at most, consistent with its slightly stronger SOC of Ni. Figure~\ref{fig:torque_eta}(b) displays the $\Gamma$ dependence for Cu/FM systems. Interestingly, Cu/Ni exhibits a non-monotonic dependence, attributable to the sharp energy dependence of $t_{yx}$ near the Fermi level [see Fig.~\ref{fig:torque}(c)]. As a result, Co yields a larger torque in the clean limit, whereas Ni produces a larger torque in the intermediate disorder regime. These results demonstrate that even for a fixed NM, the FM dependence of OT can vary, highlighting that there is no universal rule for selecting an optimal FM. Instead, the dependence varies strongly with the specific NM-FM combination and extrinsic factors such as disorder.

\begin{figure}[b]
	\center\includegraphics[width=0.48\textwidth]{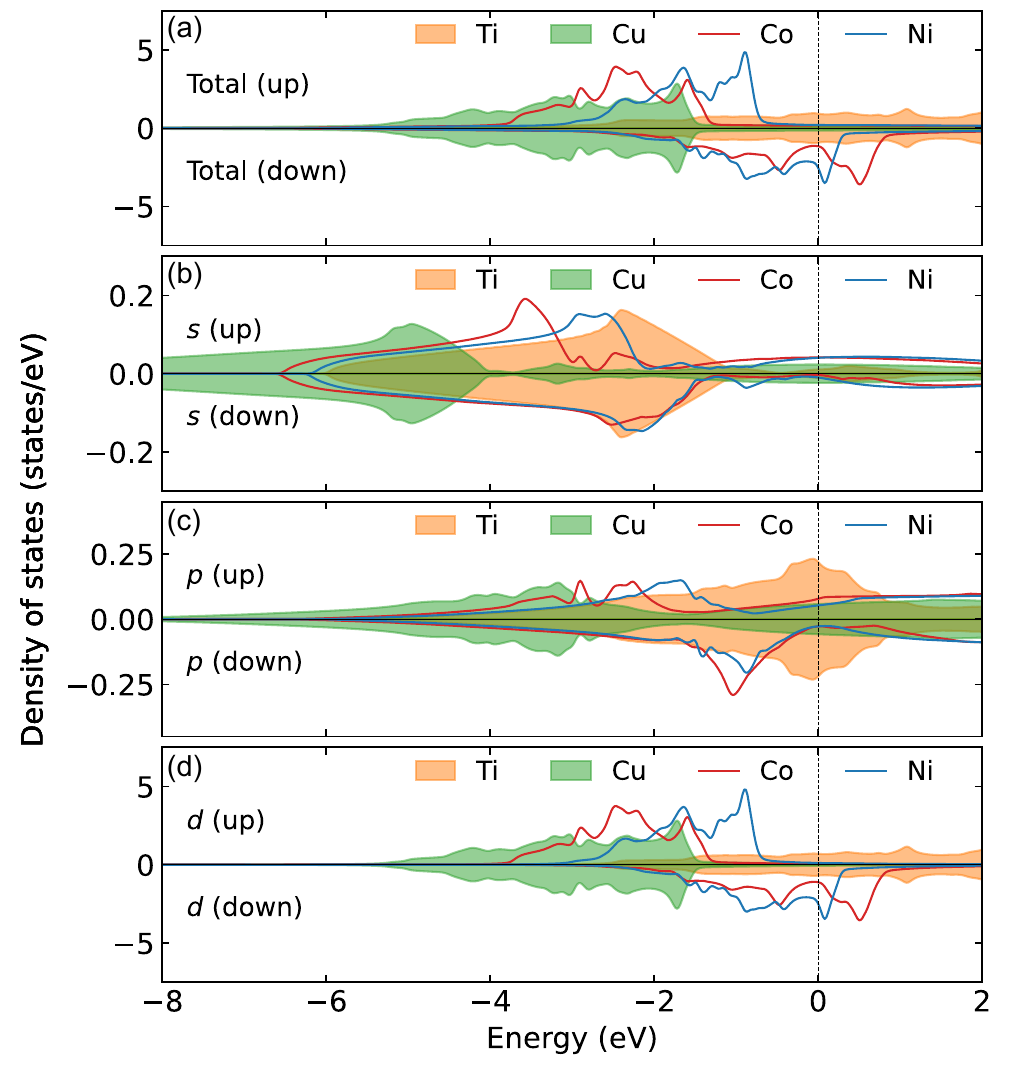}
	\caption{(a) Total DOS of bulk Ti, Cu, Co, and Ni. (b)-(d) Projected DOS for the (b) $s$, (c) $p$, and (d) $d$ orbitals. The Fermi level is set to 0~eV.}
	\label{fig:dos} 
\end{figure}

To gain insight into the effect of NM-FM matching on the OT, we compare the bulk density of states (DOS) of the constituent materials [Figs.~\ref{fig:dos}(a)-\ref{fig:dos}(d)]. The DOS of Ti shows no notable difference in its overlap with the DOS of Co versus Ni. By contrast, the total DOS of Cu overlaps more strongly with that of the spin-majority bands of Co than with that of Ni. This feature mainly originates from the $d$-orbital states [Fig.~\ref{fig:dos}(d)], which are expected to dominate the orbital response. Because the damping-like torque is obtained from a sum over occupied states [Eq.~\eqref{eq:t_ij_inter}], not only the matching at the Fermi energy but also that over an energy window below it can influence the torque. Therefore, this result suggests that the FM dependence of OT can be affected by the specific NM-FM material combination in addition to the intrinsic properties of the FM layer itself.

We further investigate the possibility of a self-induced SOT from the FM layer. To estimate this contribution, we consider hypothetical NM systems in which the $d$-orbital character is suppressed and the OHE is therefore strongly reduced. Specifically, we shift the chemical potential of Ti and Cu downward by 4~eV and 7~eV, respectively, and denote these systems as Ti$^*$ and Cu$^*$. Figures~\ref{fig:dos}(b)-\ref{fig:dos}(d) indicate that the valence electrons in Ti$^*$ and Cu$^*$ are dominated by $s$-orbital character. In Ti$^*$/FM and Cu$^*$/FM bilayers, the OHE contribution from the NM is thus expected to be negligible, while the self-induced SOT from the FM may remain. The calculated current-induced torques in these systems are shown by the dotted lines in Figs.~\ref{fig:torque}(b) and \ref{fig:torque}(c), respectively. We find that both the Ti$^*$/FM and Cu$^*$/FM bilayers exhibit negligible torque at the Fermi energy, indicating that the self-induced SOT is not a dominant contribution to the torque.

\begin{figure}[t]
	\center\includegraphics[width=0.48\textwidth]{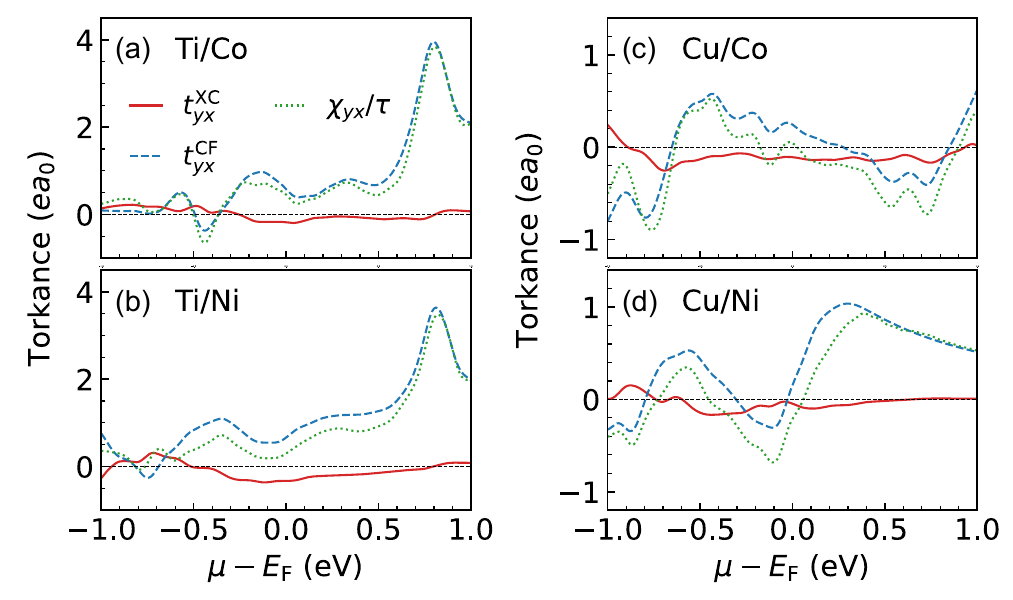}
	\caption{Calculated values for each term appearing in Eq.~\eqref{eq:continuity_approx} for (a) Ti/Co, (b) Ti/Ni, (c) Cu/Co, and (d) Cu/Ni bilayers. The energy broadening is assumed to be $\Gamma = 25$~meV.}
	\label{fig:torque_continuity} 
\end{figure}

	So far we have focused on the exchange torque [Eq.~\eqref{eq:t_xc}], which represents the transfer of angular momentum from conduction electrons to the local spin moment of the FM layer. Meanwhile, another important issue is how much the generated OAM is absorbed into the lattice. The OAM is normally expected to be largely quenched in crystals. To investigate this, we recall the continuity equation for angular momentum discussed in Sec.~\ref{sec:method_torque}. Integrating spatially Eq.~\eqref{eq:continuity} over the entire sample, the flux of the angular momentum current vanishes; thus, we obtain the following relation in terms of spatially averaged values: 
	\begin{equation}\label{eq:continuity_average}
	\mathbf{T}^\mathrm{XC}  = - \frac{\partial \mathbf{J}}{\partial t}  +\mathbf{T}^\mathrm{CF}. 
	\end{equation}
	This equation states that the exchange torque is balanced by the net accumulation rate and the crystal field torque. Note that this relation holds only at the global level, not locally. In the linear regime, the current-induced responses can be written as $T_y^\mathrm{XC/CF} = t_{yx}^\mathrm{XC/CF} E_x $ and $J_y = \chi_{yx} E_x$, which approximately yields
	\begin{equation}\label{eq:continuity_approx}
		t_{yx}^\mathrm{XC}  \approx - \frac{\chi_{yx}}{\tau}  + t_{yx}^\mathrm{CF},
	\end{equation}
	where $\tau = \hbar/\Gamma$ is the lifetime. While $t_{yx}^\mathrm{XC}$ is evaluated by Eqs.~\eqref{eq:t_ij_intra} and \eqref{eq:t_ij_inter}, $\chi_{yx}$ and $t_{yx}^\mathrm{CF}$ are obtained by replacing $\hat{T}_y^\mathrm{XC}$ with $\hat{J}_y$ and $\hat{T}_y^\mathrm{CF}$, respectively. The calculated values are shown in Figs.~\ref{fig:torque_continuity}(a)-\ref{fig:torque_continuity}(d). The results indicate that $t_{yx}^\mathrm{CF}$ is much larger than $t_{yx}^\mathrm{XC}$ and very close to $\chi_{yx}/\tau$ in general. This implies that a large portion of angular momentum is absorbed into the lattice, as typically expected, and only a relatively smaller part is transferred to the magnetization. It is worth noting that $t_{yx}^\mathrm{XC}$ is nevertheless still sizable. This also suggests the possibility of enhancing the OT by suppressing the crystal field torque, which may be of potential interest for orbitronic engineering.

\subsection{Dependence on NM thickness: bulk origin of OT}\label{sec:results_thickness}

As discussed in Sec.~\ref{sec:phenomenology}, the calculated torkance $t_{yx}$ includes not only the OT originating from the bulk orbital current in NMs but also possible contributions from the FM and the NM/FM interface. To identify the bulk NM origin in our systems, we examine the dependence of $t_{yx}$ on the NM layer thickness.

\begin{figure}[t]
	\center\includegraphics[width=0.48\textwidth]{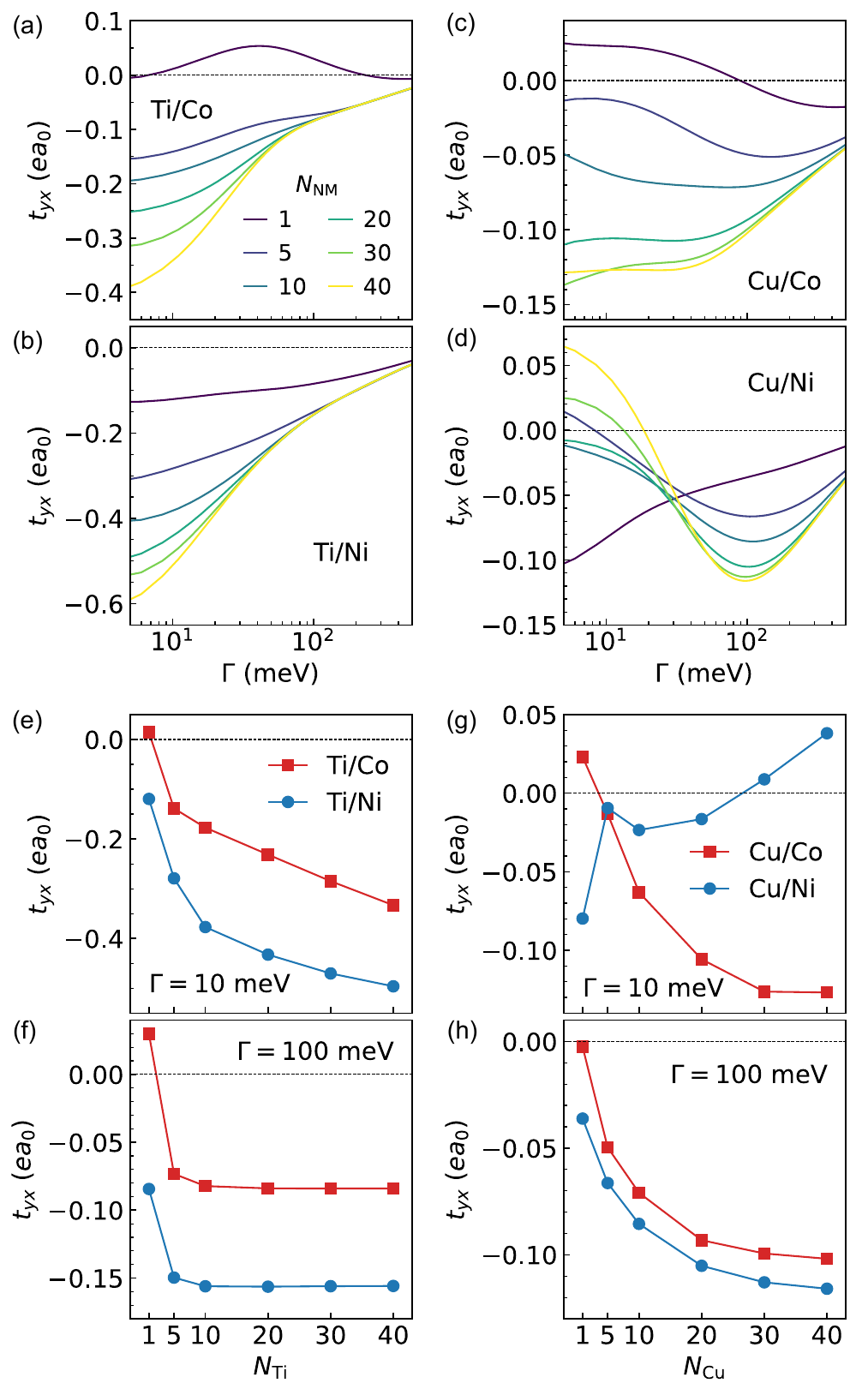}
	\caption{Torkance $t_{yx}$ as a function of the broadening energy $\Gamma$ in (a) Ti/Co, (b) Ti/Ni, (c) Cu/Co, and (d) Cu/Ni bilayers, with $N_\mathrm{NM}=1,5,10,20,30,40$ and $N_\mathrm{FM}=10$. (e),(f) $N_\mathrm{Ti}$ dependence of $t_{yx}$ in Ti/FM bilayers for $\Gamma = 10$~meV and $\Gamma = 100$~meV. (g),(h) $N_\mathrm{Cu}$ dependence of $t_{yx}$ in Cu/FM bilayers for $\Gamma = 10$~meV and $\Gamma = 100$~meV. }
	\label{fig:torque_thickness} 
\end{figure}

Figures~\ref{fig:torque_thickness}(a)-\ref{fig:torque_thickness}(d) show $t_{yx}$ as a function of the broadening energy $\Gamma$ for various NM layer numbers $N_\mathrm{NM}$, while $N_\mathrm{FM}=10$ is fixed. For $N_\mathrm{NM}=1$, the NM bulk contribution is expected to be negligible, so the torque mainly originates from the NM/FM interface. As $N_\mathrm{NM}$ increases, the bulk contribution becomes more pronounced, while the interfacial contribution would remain nearly unchanged. Comparing the results for $N_\mathrm{Ti}=1$ and $N_\mathrm{Ti}=40$ [Figs.~\ref{fig:torque_thickness}(a) and ~\ref{fig:torque_thickness}(b)], we find that the bulk Ti contribution dominates once the Ti layer becomes sufficiently thick for both Ti/Co and Ti/Ni. In particular, the thickness dependence becomes more prominent as $\Gamma$ decreases, while it weakens for $\Gamma \gtrsim 100 $~meV.

Noticeable thickness dependences of $t_{yx}$ are also observed in Cu/FM bilayers [Figs.~\ref{fig:torque_thickness}(c) and ~\ref{fig:torque_thickness}(d)]. Interestingly, for Cu/Ni, the variation of $t_{yx}$ with $N_\mathrm{Cu}$ reverses sign between clean and disordered regimes. For $\Gamma \lesssim 30$~meV, $t_{yx}$ increases in the positive direction with $N_\mathrm{Cu}$, implying that the Cu bulk contribution in this regime has the opposite sign to the OT expected for a positive orbital Hall conductivity. Consequently, the total torque becomes positive in the thick limit, unlike in other systems. This sign reversal recalls the ambiguity in the theoretically predicted sign of the orbital Hall conductivity of Cu. In addition, as discussed in Secs.~\ref{sec:method_torque} and \ref{sec:results_torque}, the conventional definition of the orbital current is not directly related to the torque acting on the magnetization. The conventional orbital Hall conductivity can serve as a good indicator only when the crystal field torque appearing in Eq.~\eqref{eq:continuity} or Eq.~\eqref{eq:continuity_average} is sufficiently small. In this regard, our results reveal the existence of a discrepancy between the theoretically calculated bulk orbital Hall conductivity and the effective orbital Hall conductivity measured in OT experiments.

To more clearly visualize the thickness dependence, we plot $t_{yx}$ as a function of $N_\mathrm{NM}$ for two selected broadening energies, $\Gamma= 10$~meV and $\Gamma=100$~meV, corresponding to the clean and disordered regimes, respectively [Figs.~\ref{fig:torque_thickness}(e)-\ref{fig:torque_thickness}(h)]. The torque is strongly suppressed for $N_\mathrm{NM}=1$, indicating a minor interfacial contribution in all cases. This aligns with experiments that report negligible damping-like torque in NM/Py devices with ultrathin ($\sim 0.5$~nm) Ti and Cu layers.~\cite{greening2020current} For Ti/FM [Fig.~\ref{fig:torque_thickness}(f)], the torque saturates rapidly, which seems to agree with recent first-principles results showing a fast decay of orbital currents within a few atomic layers.~\cite{rang2024orbital} However, we note that Ref.~\cite{rang2024orbital} investigated orbital currents injected from the FM, which differs from the OT mechanism considered here. Indeed, Fig.~\ref{fig:torque_thickness}(e) shows that the torque does not fully saturate even for $N_\mathrm{Ti}=40$ ($\approx9.6$~nm). Such a long characteristic length---well beyond atomic scales---shows qualitative agreement with experimental observations of long-range OT in Ti/FM devices.~\cite{choi2023observation, hayashi2023observation, xu2025observation, hayashi2025crystallographic, xu2025orbital}

A similarly long-range behavior is found in Cu/FM systems. As noted above, the opposite Cu bulk contribution in Cu/Ni in the clean regime is clearly visible in Fig.~\ref{fig:torque_thickness}(g). Interestingly, the characteristic length in Cu/FM systems remains relatively robust even for $\Gamma=100$~meV [Fig.~\ref{fig:torque_thickness}(h)], unlike Ti/FM, where the length shortens considerably with increasing disorder. The distinct thickness dependences between Ti/FM and Cu/FM thus warrant further investigation. Also, a more quantitative analysis would require explicit treatment of extrinsic scattering, which is beyond the scope of this work.

Finally, we emphasize that the NM bulk contribution mainly originates from the OT. Although both OT and self-induced SOT can, in principle, depend on NM thickness (see Sec.~\ref{sec:phenomenology}), the latter contribution is likely minor for the following reasons. First, when the OHE in the NM layer is suppressed, the torque becomes negligible, as shown in Figs.~\ref{fig:torque}(b) and \ref{fig:torque}(c). Second, first-principles calculations show that the spin Hall conductivity of ferromagnetic Ni is significantly large, comparable to that of Pt, whereas that of ferromagnetic Co is negligible.~\cite{go2024first} In our calculations, however, the torques in NM/Co and NM/Ni are of the same order of magnitude. Lastly, the length scale of a self-induced SOT is governed by the spin diffusion length of the NM, which can reach several hundred nanometers in light NMs such as Cu.~\cite{yakata2006temperature} By contrast, the thickness dependence observed here extends only up to about 10~nm, suggesting a much shorter relevant length scale. Meanwhile, we note that the possibility of a self-induced OT cannot be entirely ruled out. The orbital Hall conductivities for Co and Ni are comparable,~\cite{salemi2022first, go2024first} and an orbital diffusion length on the order of a few nanometers is plausible. A clear identification and separation of such effects are left for future work.

\section{Conclusion}

In this work, we examined current-induced torques in NM/FM bilayers using realistic tight-binding models constructed from first-principles electronic structures. By directly evaluating the torque in heterostructures, our approach provides quantitative results without relying on a simplified picture based on the bulk properties of individual NM and FM layers. We systematically investigated the torque depending on material combination and NM layer thickness. Our results show that both Ti/FM and Cu/FM bilayers exhibit sizable damping-like torques, ascribed to the OAM generation in the bulk NMs. The dependence on the FM choice between Co and Ni, however, differs between Ti and Cu systems: Ni yields a larger torque in Ti/FM, whereas Co produces a stronger torque in Cu/FM. This contrasting behavior indicates that the FM-dependent OT efficiency is not governed solely by the SOC strength of the FM but is also strongly affected by orbital hybridization or interfacial characteristics across the NM/FM interface. Thickness-dependent analyses further reveal that the torque predominantly arises from the NM bulk, confirming the bulk nature of the OT. The sign of the OT does not necessarily coincide with that of the bulk orbital Hall conductivity, as also indicated by a recent experiment,\cite{vijayan2025observation} calling for careful assessment of sign determination in OT measurements. The characteristic length ranges from the atomic scale to over 10 nm, depending on the disorder strength. These findings provide valuable guidance for engineering efficient OT devices based on light metals.


%
%

%

\begin{acknowledgments}
This work was supported by the Swedish Research Council (VR), the Knut and Alice Wallenberg Foundation (Grants No.\ 2022.0079 and No.\ 2023.0336), and the Wallenberg Initiative Materials Science for Sustainability (WISE) funded by the Knut and Alice Wallenberg Foundation. We further acknowledge support from the EIC Pathfinder OPEN grant No.\ 101129641 “OBELIX”. The calculations were supported by resources provided by the National Academic Infrastructure for Supercomputing in Sweden (NAISS) at NSC Link\"oping, partially funded by VR through Grant No.\ 2022-06725. 
\end{acknowledgments}

\bibliography{ref}

\end{document}